\documentclass{article}
\usepackage{graphicx} 
\usepackage{subfig}
\usepackage[hcentering, bindingoffset = 10mm, right = 20 mm, left = 10 mm, top=20 mm, bottom = 20 mm]{geometry}
\usepackage{amssymb, amsmath, amsfonts, mathtools}
\usepackage{tcolorbox}
\usepackage{xcolor}
\usepackage{wrapfig}
\usepackage{tikz}
\usepackage{hyperref}

\definecolor{linkcolor}{rgb}{0,0,1}
\definecolor{citecolor}{rgb}{0.1,0.5,0.1}
\definecolor{urlcolor}{rgb}{0.7,0.1,0.1}

\usetikzlibrary{positioning,fit,shapes.geometric}
\usetikzlibrary {decorations.pathreplacing}
\usetikzlibrary {arrows.meta}
\usepackage{ytableau}
\usepackage{myfunc}
\usepackage[natbib=true, backend = bibtex, style=numeric-comp, sorting=none]{biblatex}
\hypersetup{pdfstartview=FitH,  linkcolor=linkcolor,urlcolor=urlcolor,citecolor=citecolor, colorlinks=true}

\title{Measuring Chern-Simons level $k$ by braiding $SU(2)_k$ anyons}
\author{{ Artem Belov$^{a,c}$\thanks{e-mail: belov.artem@phystech.su }, \;  Andrey Morozov$^{a,b,c}$\thanks{e-mail: andrey.morozov@itep.ru}}
}
\date{}

\begin{document}
\ytableausetup{mathmode, boxframe=0.07em, boxsize=0.5em}
\maketitle

\vspace{-5.5cm}
\begin{center}
\hfill ITEP/TH-31/24\\
\hfill IITP/TH-26/24\\
\hfill MIPT/TH-25/24\\
\end{center}
\vspace{3.6cm}

\begin{center}

$^a$ {\small {\it NRC ``Kurchatov Institute'', Moscow 123182, Russia}}\\
$^b$ {\small {\it Institute for Information Transmission Problems, Moscow 127994, Russia}}\\
$^c$ {\small {\it Moscow Center for Advanced Studies, Kulakova str. 20, Moscow 123592, Russia}}
\end{center}

\vspace{1cm}

\begin{abstract}
Chern-Simons theory in application to the quantum computing is actively
developing at the present. However, most discussed are the questions of
using materials with known parameters and building corresponding quantum
gates and algorithms. In this paper we discuss opposite problem of
finding Chern-Simons level $k$ in the unknown material. For this purpose, we use the previously derived braiding rules for Chern-Simons $SU(2)_k$ anyons. Using certain operations (turnarounds) on three anyons, one can measure probabilities of annihilation of pairs of anyons, which depend on the parameter of the theory. Therefore, Chern-Simons level $k$ can be found from such an experiment. It is implied that anyons additionally possess certain properties which are required for topological quantum computations.
\end{abstract}

\section{Introduction}
One of the most popular areas of modern physics is quantum computing due to its possible applications in many resource-heavy problems. Many scientific groups including ones working at Google \cite{PhysRevLett133050402} or Microsoft \cite{10106312337825} are involved in researching of quantum computing and specifically topological quantum computing.  At the moment, one of the key challenges facing researchers is quantum noise, which hinders computation. Working with this problem, people have thought about using topologically invariant quantities for computation. Thus, the study of topological quantum computing is actively pursued nowadays \cite{Koh2024}, \cite{Hadjiivanov2024jho}, \cite{Xu2024}, \cite{tenHaaf2024}.

The key object in topological quantum computing is hypothetical particles called anyons. Scientific topics related to them have been actively developed recently. They were described in some quantum models \cite{Kitaev2005hzj} and fractional quantum Hall effect \cite{Fradkin1991wy}, \cite{Moore1991ks}. Anyons have unusual properties \cite{Arovas1985yb}, \cite{Frohlich1990ww} which differ from bosons and fermions. After turning them around each other, the state changes neither by $+1$ nor $-1$, but by some phase for Abelian anyons and some operators for non-Abelian anyons. This property is called fractional statistics (see more for example in \cite{Greiter}). It is argued that anyons can be produced from matter by interaction with the Chern-Simons field \cite{Arovas1985yb}, \cite{IENGO1992179}. Anyons arising from such interaction are called Chern-Simons anyons. 

It has been proposed in \cite{Kitaev1997wr} to use topological properties of anyon worldlines for quantum computation due to computational robustness to noise (see also e.g. \cite{Nayak2008zza} or \cite{Field2018} for the review). To describe such calculations it would be necessary to compute topological observables in the Chern-Simons theory - Wilson loops. However, Wilson loops in SU(N) Chern-Simons theories have been shown to be related to Jones polynomials \cite{Witten1988hf}, which can be used as a rather simple way to compute quantum gates. Quantum gates were derived for Ising and Fibonacci (see \cite{Nayak2008zza} for example), and more general types of $SU(N)$ Chern-Simons anyons \cite{KOLGANOV2023116072}, \cite{Kolganov2020}. 

Despite some physical anyon theories have been proposed and braid statistics for the $\nu=1/3$ fractional quantum Hall effect has been observed \cite{Nakamura2020mok}, it is possible that one somehow could discover solitary anyons in condensed matter, which then can possibly be adopted to topological quantum computing \cite{MELNIKOV2018491}, \cite{Mironov2024ycl}. For one who somehow get $SU(2)_k$ Chern-Simons anyons in unknown material we propose algorithm for retrieving level $k$.

\subsection{Flux-tube-charged-particle anyon example and Chern-Simons anyons}
Let us describe anyon properties using semi-physical example, proposed by Wilczek \cite{Wilczek1982wy}. Consider a thin solenoid infinite along z-axis. Assume these so called "flux-tube-charged-particle" solenoids have flux $\Phi$ and charge $q$. Then, according to Aharonov-Bohm effect, exchanging them (fig. \ref{fig:solenoid1}) gives $\exp(-i q \Phi)$ factor in the wavefunction which differs from $\pm 1$, thus producing an Abelian anyon.
\begin{figure}[h!]
  \centering
  \subfloat[Exchange of two \\ flux-tube-charged-particle]{
  \begin{tikzpicture}[scale=0.7]
        \draw[ultra thick] (-1,0)--(-1,-2);
        \draw[ultra thick] (1,0)--(1,-2);
        \fill[blue!40, semitransparent] (-4,-1)--(-2,1)--(4,1)--(2,-1)--(-4,-1);
        \draw[->, ultra thick] (-1,0)--(-1,2)node[left]{$\Phi, q \;$};
        \filldraw [black] (-1,0) circle [radius=2pt];
        \draw[->, ultra thick] (1,0)--(1,2)node[left]{$\Phi, q \;$};
        \filldraw [black] (1,0) circle [radius=2pt];
        \draw[->, black, thick] (-2.5,0.3)-- ++(-0.5,-0.5)node[right]{x};
        \draw[->, black, thick] (-2.5,0.3)-- ++(0,0.7)node[left]{z};
        \draw[->, black, thick] (-2.5,0.3)-- ++(0.7,0)node[below]{y};
        \draw[->, black, thick] (-0.9,0.1) .. controls (-0.3,0.5) and (0.7, 0.5) .. (0.9,0.1);
        \draw[->, black, thick] (0.9, -0.1) .. controls (0.3, -0.5) and (-0.7, -0.5) .. (-0.9,-0.1);
    
    \end{tikzpicture}
  \label{fig:solenoid1}}
  \hfill
  \subfloat[Fusion diagram for two \\ flux-tube-charged-particle]{
  \begin{tikzpicture}
    \draw[-Stealth, blue, thick] (-1,-1)node[left]{$(\Phi_1 , q_1)$}--(-0.5, -0.5);
    \draw[blue, thick](0,0)--(-0.6,-0.6);
    
    \draw[-Stealth, blue, thick] (1,-1)node[right]{$(\Phi_2 , q_2)$}--(0.5, -0.5);
    \draw[blue, thick](0,0)--(0.6,-0.6);
    
    \draw[-Stealth, blue, thick] (0,0)--(0, 0.5);
    \draw[blue, thick](0,0.4)--(0,1)node[above]{$(\Phi_1 + \Phi_2 , q_1 + q_2)$};
    
    \end{tikzpicture}
  \label{fig:solenoid2}}
  \hfill
  \subfloat[Fusion into $\varnothing$ (zero charge and flux) for two flux-tubes-charged-particles]{
  \begin{tikzpicture}
    \draw[-Stealth, blue, thick] (-1,-1)node[left]{$(\Phi , q)$}--(-0.5, -0.5);
    \draw[blue, thick](0,0)--(-0.6,-0.6);
    
    \draw[-Stealth, blue, thick] (1,-1)node[right]{$(-\Phi , -q)$}--(0.5, -0.5);
    \draw[blue, thick](0,0)--(0.6,-0.6);
    
    \draw[dashed, blue, thick] (0,0) -- (0, 1) node[midway, right]{$\varnothing$}node[above]{$(0,0)$};
    
    \end{tikzpicture}
  \label{fig:solenoid3}}
  \caption{}
\end{figure}
In algebraic description, this gives braiding rules. One also can fuse two solenoids, to get another one (fig. \ref{fig:solenoid2}). After fusing solenoids, their charges and fluxes are summed up. In algebraic description, it corresponds to the fusion rules. Sometimes fusion of a pair of anyons can produce trivial representation $\varnothing$. Considering flux-tubes-charged-particles, this means $\Phi=0$ and $q=0$ (fig. \ref{fig:solenoid3}). Turning any anyon around $\varnothing$ changes nothing (wavefunction does not change). That is why $\varnothing$ can be interpreted as a vacuum. When one considers fusion rules, it is often convenient to use diagrams such as on figure \ref{fig:solenoid2} and \ref{fig:solenoid3}.

Let us now talk about Chern-Simons anyons. Such anyons, as previously said, can be obtained by ordinary particles interacting with Chern-Simons field \cite{Arovas1985yb}. The Chern-Simons action looks like:
\begin{equation}
    S_\text{CS}[\mathcal{A}] = \frac{k}{4\pi} \int \text{Tr}[\mathcal{A}\wedge d\mathcal{A} + \frac{2}{3} \mathcal{A}\wedge \mathcal{A}\wedge \mathcal{A}],
\end{equation}
where $\mathcal{A}=\mathcal{A}_\mu dx^\mu$ is Chern-Simons gauge field and $k$ is called level of the theory. Topological quantum computations are provided by using topological invariant observables. In Chern-Simons theory, the average of Wilson loop
\begin{equation}
    \langle W \rangle = \left\langle \text{Tr Pexp} \oint \mathcal{A} \right\rangle
\end{equation}
is such an observable. These observables are formed by worldlines, which can form knots. It is argued that fractional statistics can be realised only in 3-dimensional space-time (2 spatial and 1 time dimension). Fortunately, for topological quantum computing one does not need to calculate the integral, since it has been proven that such integrals are Jones polynomials \cite{Witten1988hf} when gauge group is $SU(2)$.

\subsection{What is this paper about}\label{sec:whatabout}

\begin{wrapfigure}[10]{r}{0.45\textwidth}
  \begin{center}
  \vspace{-2cm}
    \subfloat[Annihilation probability $p(d,t)$ in $\varnothing$ representation]{
    \begin{tikzpicture}[scale=0.45, decoration=brace]
        \draw [ultra thick, ->] (-3,-1) -- (-3,8)node[left]{Time\;};
        \draw[densely dashed, blue, thick] (0,-1)--(0,0);
        \draw[blue, thick] (0,0) .. controls (0,0.5) and (-1.5,1) .. (-1.5,1.5);
        \draw[blue, thick] (0,0) .. controls (0,0.5) and (1.5,1) .. (1.5,1.5);
        \draw[blue, thick] (-1.5, 1.5) .. controls (-1.5, 2) and (-1, 2.5) .. (-1, 3);
        \draw[blue, thick] (1.5, 1.5) .. controls (1.5, 2) and (1, 2.5) .. (1, 3);
        \draw[<->,thick] (0.9,3)--(-0.9,3)node[above, midway]{$d$};
        \draw[blue, thick] (1,3)--(1,6.5);
        \draw[blue, thick] (-1,3)--(-1,6.5);
        \draw[decorate,thick] (-1.1, 6.8)--(1.1, 6.8) node[above, midway, align=left]{ $p(d,t)$};
    \end{tikzpicture}
    \label{fig:desc1}
    }
    \hfill
    \subfloat[Annihilation probability $\mathcal{P}_n$ after $n$ turnarounds]{
    \begin{tikzpicture}[scale=0.45, decoration=brace]
        \draw[->, ultra thick] (-1,0)--(-1,9)node[left]{Time\;};
        \draw[blue, thick] (0,0)--(0,1);
        \newanyon(1,1);
        \rcross(0,1);
        \rcross(0,2);
        \fill[blue] (0.5,3.5) circle [radius=1.5pt];
        \fill[blue] (0.5,4) circle [radius=1.5pt];
        \fill[blue] (0.5,4.5) circle [radius=1.5pt];
        \rcross(0,5);
        \rcross(0,6);
        \draw[blue, thick] (0,7)--(0,9);
        \draw[blue, thick] (2,1)--(2,7.5);
        \draw[blue, thick] (1,7)--(1,7.5);
        \draw[decorate,thick] (0.9, 7.8)--(2.1, 7.8) node[above, at end, align=left]{ $\mathcal{P}_n(k,d,t)$};
    \end{tikzpicture}
    \label{fig:desc2}
    }
  \end{center}
  \caption{}
\end{wrapfigure}
Solitary anyons could be used for topological quantum computations (see \cite{KOLGANOV2023116072}, \cite{article} or \cite{Xu2024} for example). However there are some requirements to make this possible:
\begin{enumerate}
    \item one has to be able to obtain a pair of anyons from a vacuum (fig. \ref{fig:desc1} or right part of \ref{fig:desc2}). It is reasonable to assume that anyon pair obtained from vacuum is in $\varnothing$ representation;
    \item  one has to be able to control the position of anyons and to move them according to the required algorithm;
    \item a pair of anyons in the representation $\varnothing$, brought close enough has to annihilate after a long enough time.
\end{enumerate}
We assume that these requirements are satisfied for the anyons studied in the present paper.

Here and further under the $n$ turnarounds the following is meant: there is one anyon (the first), which was somehow obtained in advance, a pair of anyons (the second and the third) was obtained from vacuum, then $n$ double-twists (turnarounds) of the first and the second anyon were made (see fig. \ref{fig:desc2}). Acting like this $n$ turnaround operator we denote $\hat{\mathcal{T}}^n$ during this article.

Moreover, throughout this article by $\mathbb{P}_n$ we mean probability of a pair of measurable anyons (right pair) to remain in the $\varnothing$ representation after $n$ turnarounds. By $\mathcal{P}_n$ we mean annihilation probability for a pair of measurable anyons (right pair) after $n$ turnarounds. To connect $\mathbb{P}_n$ and $\mathcal{P}_n$ one have to beforehand measure (experimentally) annihilation probability $p(d,t)$ for an anyon pair in $\varnothing$ representation (fig. \ref{fig:desc1}). In section \ref{sec:alg} we derive probability $\mathbb{P}_n$. However, the experimenter, when measuring, will observe not representations but rather annihilation or its absence. Consequently, during experiment one will get probability $\mathcal{P}_n$, which is not the same as $\mathbb{P}_n$ since it is possible that the pair of anyons in representation $\varnothing$ did not have enough time to annihilate (as we will discuss in section \ref{sec:expdesc} using fig. \ref{fig:possible2}).  That is why to connect $\mathbb{P}_n$ and $\mathcal{P}_n$ together, one should measure probability of anyon pair in $\varnothing$ representation to annihilate, depending on distance $d$ and waiting time $t$ before the main part of experiment (fig. \ref{fig:desc2}). The connection between $\mathbb{P}_n$ and $\mathcal{P}_n$ is as follows
\begin{equation}\label{eq:porbprobprob}
    \mathbb{P}_n(k) = \frac{\mathcal{P}_n(k,d,t)}{p(d,t)},
\end{equation}
where $\mathcal{P}_n(k,d,t)$ and $p(d,t)$ are measured experimentally, and $\mathbb{P}_n(k)$ should be compared with algebraically derived $\mathbb{P}(k)$ in section \ref{sec:alg}. So the algorithm for retrieving $k$ as follows:
\begin{enumerate}
    \item experimentally measure the probability $p(d,t)$ of anyon pair in $\varnothing$ representation to annihilate. To do this, one has to get a pair of anyons from a vacuum and place them at a distance $d$ (fig. \ref{fig:desc1});
    \item for $n$ turnarounds measure annihilation probability $\mathcal{P}_n(k,d,t)$ (fig. \ref{fig:desc2}) and then use formula \eqref{eq:porbprobprob} to get $\mathbb{P}_n(k)$. Start with $n=1$ turnarounds and take the next one from the list $(1,3,9,27,...,3^\#)$ every time when $\mathbb{P}_n(k)$ is definitely higher than $\mathbb{P}_n(k=24n-2) \approx 0.8$;
    \item when $\mathbb{P}_n(k)$ is somewhere below than $\mathbb{P}_n(k=24n-2) \approx 0.8$, then one should repeat measurements until confidence interval covers $\mathbb{P}_n(k)$ for only one $k$.
\end{enumerate}

\subsubsection*{Definitions}
Let us collect in one place all the notations we are going to use with no description further:
\begin{itemize}
    \item $\mathbb{P}_n(k)$ is the probability of the measurable pair of anyons (right pair) to remain in the $\varnothing$ representation after $n$ turnarounds;
    \item $\mathcal{P}_n(k,d,t)$ is the annihilation probability for the measurable pair of anyons (right pair) after $n$ turnarounds;
    \item $p(d,t)$ is the probability of anyon pair in $\varnothing$ representation to annihilate.
\end{itemize}

\section{Anyons, representations and their worldlines}
\subsection{$SU(2)$ representations and anyons}
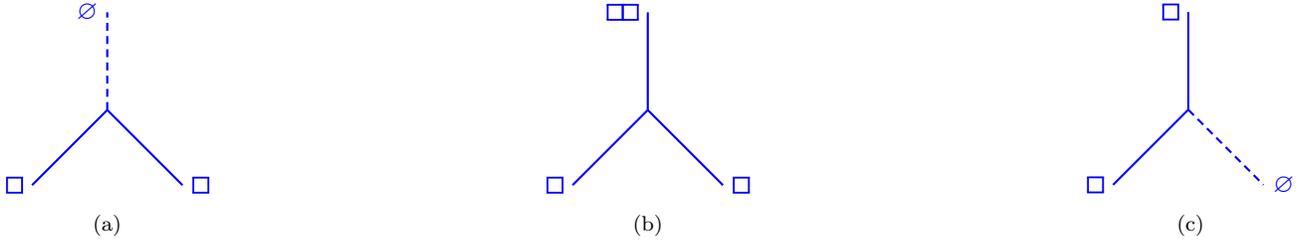
\begin{figure}[h!]
  \begin{center}
    \subfloat[]{
    \begin{tikzpicture}[scale=1, decoration=brace]
       \draw[blue, thick, densely dashed] (0,0)--(0,1.3)node[above, left]{$\varnothing$};
       \draw[blue, thick] (0,0)--(1,-1)node[right]{\ydiagram{1}};
       \draw[blue, thick] (0,0)--(-1,-1)node[left]{\ydiagram{1}};
    \end{tikzpicture}
    \label{fig:fus0}
    }
    \hfill
    \subfloat[]{
    \begin{tikzpicture}[scale=1, decoration=brace]
        \draw[blue, thick] (0,0)--(0,1.3)node[above, left]{$\ydiagram{2}$};
       \draw[blue, thick] (0,0)--(1,-1)node[right]{\ydiagram{1}};
       \draw[blue, thick] (0,0)--(-1,-1)node[left]{\ydiagram{1}};
    \end{tikzpicture}
    \label{fig:fus2}
    }
    \hfill
    \subfloat[]{
    \begin{tikzpicture}[scale=1, decoration=brace]
        \draw[blue, thick] (0,0)--(0,1.3)node[above, left]{\ydiagram{1}};
       \draw[blue, thick, densely dashed] (0,0)--(1,-1)node[right]{$\varnothing$};
       \draw[blue, thick] (0,0)--(-1,-1)node[left]{\ydiagram{1}};
    \end{tikzpicture}
    \label{fig:fuswith0}
    }
  \end{center}
  \caption{Fusion diagrams for $\ydiagram{1}$ and $\varnothing$ Chern-Simons $SU(2)_k$ anyons}
\end{figure}

In the present paper we discuss Chern-Simons theory with gauge group $SU(2)$. Anyons in this case correspond to irreducible representations of quantum $SU(2)_k$ \cite{Lerda1992yg}. You can see more about quantum groups in \cite{klimyk2012quantum}. A pair of  $SU(2)_k$ Chern-Simons anyons in fundamental representation could be in $\varnothing$ representation (fig. \ref{fig:fus0}) or in $\ydiagram{2}$ representation (fig. \ref{fig:fus2})  since in $SU(2)$:
\begin{equation}\label{eq:su2FR}
    \ydiagram{1}\otimes \ydiagram{1} = \ydiagram{2} \oplus \varnothing.
\end{equation}
Usually $\ydiagram{2}$ and $\varnothing$ are also called anyons, which are different from $\ydiagram{1}$ anyons (these representations are also sometimes called topological charges). 
There is a slight difference here from what we saw on figure \ref{fig:solenoid2}: the fusion of anyon pair can proceed in different ways. Moreover, braid operators in this case constitute a non-commutative set of operators. Thus here we deal with non-Abelian anyons. Formula \eqref{eq:su2FR} also means that anyon $\ydiagram{1}$ and anti-anyon $\ydiagram{1}$ in $SU(2)$ case group is the same, since we have $\ydiagram{1}\otimes\ydiagram{1}\to\varnothing$ fusion rule. Braiding rules provide the so-called R-matrix $\sigma$ acting on tensor product of $SU(2)_k$ representations. Anyon pair sometimes can change representation, for example braid operator $\sigma$ acting on the second and the third anyon in $(\ydiagram{1}\otimes\ydiagram{1})\otimes\ydiagram{1}$ basis could change the fusion of the first and the second anyon pair. However, the block-diagonal structure of $\sigma$ keeps the system in the same representation of the tensor product. Also, from the algebraic point of view, $\varnothing$ is a trivial, one-dimensional representation. Therefore its tensor product with any representation keeps this representation and does not produce anything else (fig. \ref{fig:fuswith0}). For example
\begin{equation}
    \varnothing \otimes \ydiagram{1}=\ydiagram{1}\otimes \varnothing = \ydiagram{1} \;.
\end{equation}

\subsection{Topology and worldlines}
\begin{figure}[h!]
  \centering
  \subfloat[Despite the theory being topological, anyon system in $\varnothing$ representation can be moved through other anyons (with fixed ends) since it has no effect on the system]{
  \begin{tikzpicture}
    \draw[draw=white, double=blue, ultra thick] (-1,0)node[below, left, align=right]{anyon \\(or system of anyons)} -- (-1,1);
        \draw[draw=white, double=blue, densely dashed, ultra thick] (0,0)node[below, right]{$\varnothing$} .. controls (-2,0.75) and (-2,1.25) .. (0,2);
        \draw[draw=white, double=blue, ultra thick] (-1,1) -- (-1,2);
        \draw[ blue] (2,0)--(2,2);
        \draw[densely dashed, blue, thick] (3,0)--(3,2);
        \draw[<->,double] (0.5,1)--(1.5,1);
    \end{tikzpicture}
  \label{fig:worldlines1}}
  \hfill
  \subfloat[Anyon system in $\ydiagram{2}$ representation can not be homotopically moved (with fixed ends) into right picture (for all anyon system, which representation differs from $\varnothing$)]{
  \begin{tikzpicture}[decoration=brace]
        \draw[draw=white, double=blue, ultra thick] (-1,0)node[below, left, align=right]{anyon \\(or system of anyons)} -- (-1,1);
        \draw[draw=white, double=blue, ultra thick] (0,0)node[below, right]{$\ydiagram{2}$} .. controls (-2,0.75) and (-2,1.25) .. (0,2);
        \draw[draw=white, double=blue, ultra thick] (-1,1) -- (-1,2);
        \draw[ blue] (2,0)--(2,2);
        \draw[ blue] (3,0)--(3,2);
        \draw[<->,double] (0.5,1)--(1.5,1);
        \draw[red,thick] (0.75,0.75)--(1.25,1.25);
        \draw[red,thick] (0.75,1.25)--(1.25,0.75);
    \end{tikzpicture}
  \label{fig:worldlines2}}
\caption{Comparison of worldlines topological properties}
\end{figure}
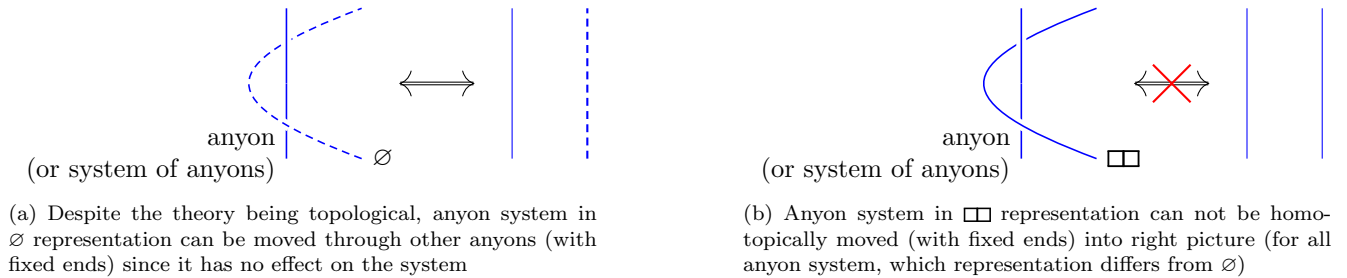

Let us now describe what does $\varnothing$ representation mean for the worldlines topology. Turning any anyons around $\varnothing$ does not change the wavefunction. Since here we are considering topological properties, we can not homotopically move worldlines with fixed end points through each other (see fig. \ref{fig:worldlines2} for example), however for $\varnothing$ representation worldline we can (see fig. \ref{fig:worldlines1}). Thus we consider $\varnothing$ representation as a vacuum. Picture \ref{fig:worldlines2} means no one can homotopically move worldline with two turnarounds into three, for example. Therefore, for different numbers $n$ of turnarounds one has different operators $\hat{\mathcal{T}}^n$. Thus, after different numbers $n$ of turnarounds there is different probability $\mathbb{P}_n(k)$ (also different annihilation probability $\mathcal{P}_n(k,d,t)$).

\section{Experiment description}\label{sec:expdesc}
\begin{wrapfigure}[11]{r}{0.45\textwidth}
  \begin{center}
  \vspace{-2cm}
    \subfloat[Measuring annihilation probability $p(d,t)$ in $\varnothing$ representation]{
    \begin{tikzpicture}[scale=0.45, decoration=brace]
        \draw [ultra thick, ->] (-3,-1) -- (-3,8)node[left]{Time\;};
        \draw[densely dashed, blue, thick] (0,-1)--(0,0);
        \draw[blue, thick] (0,0) .. controls (0,0.5) and (-1.5,1) .. (-1.5,1.5);
        \draw[blue, thick] (0,0) .. controls (0,0.5) and (1.5,1) .. (1.5,1.5);
        \draw[blue, thick] (-1.5, 1.5) .. controls (-1.5, 2) and (-1, 2.5) .. (-1, 3);
        \draw[blue, thick] (1.5, 1.5) .. controls (1.5, 2) and (1, 2.5) .. (1, 3);
        \draw[<->,thick] (0.9,3)--(-0.9,3)node[above, midway]{$d$};
        \draw[blue, thick] (1,3)--(1,6.5);
        \draw[blue, thick] (-1,3)--(-1,6.5);
        \draw[decorate,thick] (-1.1, 6.8)--(1.1, 6.8) node[above, midway, align=left]{ $p(d,t)$};
    \end{tikzpicture}
    \label{fig:desc1v2}
    }
    \hfill
    \subfloat[Measuring annihilation probability $\mathcal{P}_n$ after $n$ turnarounds]{
    \begin{tikzpicture}[scale=0.45, decoration=brace]
        \draw[->, ultra thick] (-1,0)--(-1,9)node[left]{Time\;};
        \draw[blue, thick] (0,0)--(0,1);
        \newanyon(1,1);
        \rcross(0,1);
        \rcross(0,2);
        \fill[blue] (0.5,3.5) circle [radius=1.5pt];
        \fill[blue] (0.5,4) circle [radius=1.5pt];
        \fill[blue] (0.5,4.5) circle [radius=1.5pt];
        \rcross(0,5);
        \rcross(0,6);
        \draw[blue, thick] (0,7)--(0,9);
        \draw[blue, thick] (2,1)--(2,7.5);
        \draw[blue, thick] (1,7)--(1,7.5);
        \draw[decorate,thick] (0.9, 7.8)--(2.1, 7.8) node[above, midway, align=left]{ $\mathcal{P}_n(d,t)$};
    \end{tikzpicture}
    \label{fig:desc2v2}
    }
  \end{center}
  \caption{}
\end{wrapfigure}
In figures \ref{fig:desc1v2} and \ref{fig:desc2v2}, we have introduced the basic operations of our algorithm. The probability $p(d,t)$ has to be measured beforehand (fig. \ref{fig:desc1v2}), then experiments with different number of turnarounds are performed (fig. \ref{fig:desc2v2}), and $k$ to be retrieved based on the results. We'll write more about experiment here. 

Since it is proposed that anyons in $\varnothing$ representation should annihilate, the annihilation probability $p(d,t)$ as a function of the distance and time should be measured. Consider the following experiment: one (somehow) produce anyon pair from vacuum (that means that they are in $\varnothing$ representation) then bring them together to the distance $d$ and wait for time $t$. The result is either pair of anyons annihilate or not. 

\begin{figure}[h!]
    \centering
    \subfloat[Anyons in $\varnothing$ representation at the distance $d$ can annihilate after time $t$ with $p(d,t)$ probability]{
        \begin{tikzpicture}[decoration=brace]
            \draw [ultra thick, ->] (-3,0) -- (-3,4)node[left]{Time\;};
            \draw[blue, thick](0,0)--(0,1);
            \draw[blue, thick](-1,0)--(-1,1);
            \draw[decorate,thick] (0.1,-0.1)--(-1.1,-0.1) node[midway, below]{$\varnothing$};
            \draw[blue, thick] (-1,1) .. controls (-1,1.2) and (-0.8, 1.4) .. (-0.8, 1.6);
            \draw[blue, thick] (0, 1) .. controls (0, 1.2) and (-0.2, 1.4) .. (-0.2, 1.6);
            \draw[thick, blue] (-0.2, 1.6) -- (-0.2, 2);
            \draw[thick, blue] (-0.8, 1.6) -- (-0.8, 2);
            \draw[thick, blue] (-0.2, 2) .. controls (-0.2, 2.2) and (-0.5, 2.4) .. (-0.5, 2.5) ;
            \draw[thick, blue] (-0.8, 2) .. controls (-0.8, 2.2) and (-0.5, 2.4) .. (-0.5, 2.5) ;
            \draw[<->,thick] (-0.8, 1.4)--(-0.2, 1.4) node[below, midway]{$d$};
            \draw[decorate,thick] (0.5, 4)--(0.5, 1.4) node[midway, right]{$t$};
        \end{tikzpicture}
    \label{fig:possible1}}
    \hfill
    \subfloat[Anyons in $\varnothing$ representation at the distance $d$ may not to annihilate after time $t$ with $1-p(d,t)$ probability]{
        \begin{tikzpicture}[decoration=brace]
            \draw [ultra thick, ->] (-3,0) -- (-3,4)node[left]{Time\;};
            \draw[blue, thick](0,0)--(0,1);
            \draw[blue, thick](-1,0)--(-1,1);
            \draw[decorate,thick] (0.1,-0.1)--(-1.1,-0.1) node[midway, below]{$\varnothing$};
            \draw[blue, thick] (-1,1) .. controls (-1,1.2) and (-0.8, 1.4) .. (-0.8, 1.6);
            \draw[blue, thick] (0, 1) .. controls (0, 1.2) and (-0.2, 1.4) .. (-0.2, 1.6);
            \draw[thick, blue] (-0.2, 1.6) -- (-0.2, 4);
            \draw[thick, blue] (-0.8, 1.6) -- (-0.8, 4);
            \draw[<->,thick] (-0.8, 1.4)--(-0.2, 1.4) node[below, midway]{$d$};
            \draw[decorate,thick] (0.5, 4)--(0.5, 1.4) node[midway, right]{$t$};
        \end{tikzpicture}
    \label{fig:possible2}}
    \hfill
    \subfloat[Anyons in $\ydiagram{2}$ representation can not annihilate]{
        \begin{tikzpicture}[decoration=brace]
            \draw [ultra thick, ->] (-3,0) -- (-3,4)node[left]{Time\;};
            \draw[blue, thick](0,0)--(0,1);
            \draw[blue, thick](-1,0)--(-1,1);
            \draw[decorate,thick] (0.1,-0.1)--(-1.1,-0.1) node[midway, below]{$\ydiagram{2}$};
            \draw[blue, thick] (-1,1) .. controls (-1,1.2) and (-0.8, 1.4) .. (-0.8, 1.6);
            \draw[blue, thick] (0, 1) .. controls (0, 1.2) and (-0.2, 1.4) .. (-0.2, 1.6);
            \draw[thick, blue] (-0.2, 1.6) -- (-0.2, 4);
            \draw[thick, blue] (-0.8, 1.6) -- (-0.8, 4);
        \end{tikzpicture}
    \label{fig:possible3}}
    \caption{Connection between possible experimental outcomes and anyon pair representations}
\end{figure}
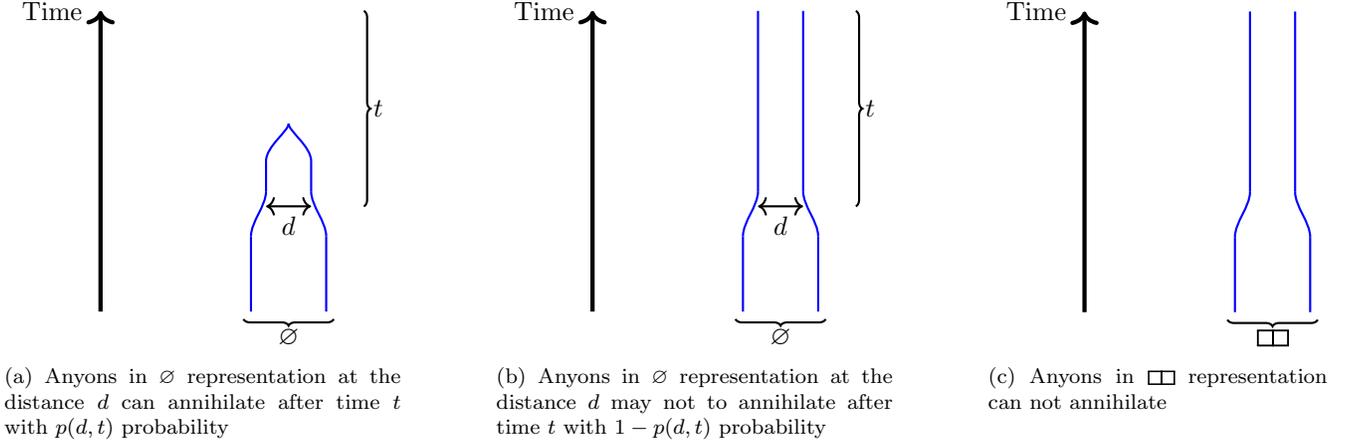

After $p(d,t)$ was measured, the preparation for the main experiment was done. During the main experiment (fig. \ref{fig:desc2v2}) three situations are possible after $n$ turnarounds: 
\begin{enumerate}
    \item with probability $\mathbb{P}_n(k) p(d,t)$ one get annihilated anyon pair in $\varnothing$ representation after measurements (fig. \ref{fig:possible1});
    \item with probability $\mathbb{P}_n(k)(1-p(d,t))$ one get non-annihilated anyon pair in $\varnothing$ representation after measurements (fig. \ref{fig:possible2});
    \item with probability $1-\mathbb{P}_n(k)$ one get non-annihilated anyon pair in $\ydiagram{2}$ representation after measurements (fig. \ref{fig:possible3}).
\end{enumerate}

As one can see the measured annihilation probability $\mathcal{P}_n(k,d,t)$ can not be compared with $\mathbb{P}_n(k)$. To get annihilation probability one should multiply probability $\mathbb{P}_n(k)$ by probability of annihilation in $\varnothing$ representation $p(d,t)$ as follows:

\begin{equation}
    \mathcal{P}_n(k,d,t)=p(d,t)\mathbb{P}_n(k).
\end{equation}
After $p(d,t)$ and $\mathcal{P}_n(k,d,t)$ was measured, one can get
\begin{equation}
    \mathbb{P}_n(k) = \frac{\mathcal{P}_n(k,d,t)}{p(d,t)},
\end{equation}
and then compare it with $\mathbb{P}_n(k)$, algebraically derived in section \ref{sec:algk3}, to retrieve $k$.

\section{Probability $\mathbb{P}_n(k)$ algebraic computations}\label{sec:alg}
\begin{wrapfigure}[14]{l}{0.3\textwidth}
    \centering
    \vspace{-0cm}
    \begin{tikzpicture}
        \draw [ultra thick, ->] (-1,0) -- (-1,4)node[left]{Time\;};
        \draw [thick,blue] (1,0) -- (1,1); 
        \draw [thick,blue] (2,0) -- (2,1);
        \rcross(1,1);
        \rcross(1,2);
        \draw [thick,blue] (1,3) -- (1,4); 
        \draw [thick,blue] (2,3) -- (2,4);
        \draw [line width=0.8pt,blue] (3,0) -- (3,2); 
        \draw [thick,blue] (3,2) -- (3,4); 
    \end{tikzpicture}
    \caption{One turnaround for the \\ $1^\text{st}$ and the $2^\text{nd}$ anyons }
    \label{fig:turnaround-operator}
\end{wrapfigure}
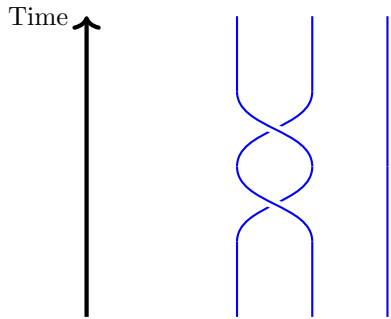
Let's compute the probability $\mathbb{P}_n$ that a pair of anyons remain in the $\varnothing$ representation after $n$ turnarounds. From the algebraic point of view, three anyons $\ydiagram{1}\otimes\ydiagram{1}\otimes\ydiagram{1}$ have the following fusion rules 
\begin{equation}
\ydiagram{1}\otimes\ydiagram{1}\otimes\ydiagram{1} = \ydiagram{3}\oplus2 \; \ydiagram{1} \;.
\end{equation}

Since two of them are in $\varnothing$ representation in the beginning, this means the whole system is in the $\ydiagram{1}\otimes \varnothing=\ydiagram{1}$ representation. Since all operators are block-diagonal, all actions on anyons keep them in $\ydiagram{1}$ representation. There are two basis vectors in $\ydiagram{1}\otimes\ydiagram{1}\otimes\ydiagram{1} \to \ydiagram{1} \;$:

\begin{equation}\label{eq:basis}
| \varnothing\rangle = \left[
\begin{tikzpicture}[baseline={([yshift=-.4ex]current bounding box.center)}, thick,scale=0.4, every node/.style={scale=1.0}]
    \draw[blue] (2,-2)--(0,0)node[anchor=south]{$\ydiagram{1}$};
    \draw[blue] (3,-1)--(2,0)node[anchor=south]{$\ydiagram{1}$};
     \draw[blue] (3,-1)--(4,0)node[anchor=south]{$\ydiagram{1}$};
     \draw[blue, densely dashed] (2,-2)--(3,-1)node[label={[label distance=0]270:$\varnothing$}]{};
     \draw[blue] (2,-2)--(3,-3)node[anchor=north]{$\ydiagram{1}$};
\end{tikzpicture}
\right], \qquad 
|\; \ydiagram{2} \; \rangle = \left[\diagThreeRight(\ydiagram{1},\ydiagram{1},\ydiagram{1},\ydiagram{1},\ydiagram{2})\right].
\end{equation}

Applying braid operator for the first and the second anyons twice $\hat{\sigma}^2$, we get turnaround operator $\hat{\mathcal{T}}$ (fig.\ref{fig:turnaround-operator}).

\subsection{$k>3$}\label{sec:algk3}
For $k>3$ the braid operator could be calculated using Racah matrices and R-matrices eigenvalues from \cite{Dhara2019} or \cite{Anokhina2014}. Here and further the so-called quantum numbers $[m]_q$ are used:
\begin{equation}
    [m]_q=\frac{q^m-q^{-m}}{q-q^{-1}},
\end{equation}
where $q=\exp(2\pi i/(2+k))$. Braid operator in \eqref{eq:basis} basis is equal to
\begin{equation}
\hat{\sigma} = \frac{1}{[2]^2}
\begin{pmatrix}
-1 & \sqrt{[3]} \\
\sqrt{[3]} & 1
\end{pmatrix}
\begin{pmatrix}
-q^{-1} & 0 \\
0 & q
\end{pmatrix} 
\begin{pmatrix}
-1 & \sqrt{[3]} \\
\sqrt{[3]} & 1
\end{pmatrix}.
\end{equation}
Let us mention that Racah matrices are unitary and real for $k>3$. In this case
\begin{equation}
\frac{1}{[2]^2}
\begin{pmatrix}
-1 & \sqrt{[3]} \\
\sqrt{[3]} & 1
\end{pmatrix}
\begin{pmatrix}
-1 & \sqrt{[3]} \\
\sqrt{[3]} & 1
\end{pmatrix}
=
\begin{pmatrix}
    1 & 0 \\
    0 & 1
\end{pmatrix},
\end{equation}
consequently turnaround operator is equal to
\begin{equation}
\hat{\mathcal{T}}=\hat{\sigma}^{2} = \frac{1}{[2]^2}
\begin{pmatrix}
-1 & \sqrt{[3]} \\
\sqrt{[3]} & 1
\end{pmatrix}
\begin{pmatrix}
q^{-2} & 0 \\
0 & q^{2}
\end{pmatrix} 
\begin{pmatrix}
-1 & \sqrt{[3]} \\
\sqrt{[3]} & 1
\end{pmatrix}
\end{equation}
and $n$ turnarounds operator is equal to
\begin{equation}
\hat{\mathcal{T}}^{n} = \frac{1}{[2]^2}
\begin{pmatrix}
-1 & \sqrt{[3]} \\
\sqrt{[3]} & 1
\end{pmatrix}
\begin{pmatrix}
q^{-2n} & 0 \\
0 & q^{2n}
\end{pmatrix} 
\begin{pmatrix}
-1 & \sqrt{[3]} \\
\sqrt{[3]} & 1
\end{pmatrix}
=
\frac{1}{[2]^2}
\begin{pmatrix}
q^{-2n}+\sqrt{[3]}q^{2n} & \bullet \\
\bullet & q^{2n}+\sqrt{[3]}q^{-2n}
\end{pmatrix}
.
\end{equation}
Probability amplitude that the pair of anyons remains in the same representation is given by diagonal elements. The probability to change the representation is $1-\mathbb{P}_n$, thus here we only give the diagonal elements that correspond to the amplitude of the probability $\mathbb{P}_n$. Since absolute values of diagonal elements are equal:
\begin{equation}
    \left|\frac{q^{-2n}+[3]q^{2n}}{[2]^2} \right|=\left|\frac{q^{2n}+[3]q^{-2n}}{[2]^2} \right|
\end{equation}
the probability $\mathbb{P}_n$ of two right anyons preserve representation ($|\varnothing\rangle\to|\varnothing\rangle$or$|\,\ydiagram{2}\,\rangle\to|\,\ydiagram{2}\,\rangle$) after $n$ turnarounds is equal to
\begin{equation}\label{eq:main}
    \mathbb{P}_n(k) = \left|\frac{q^{2n}+[3]q^{-2n}}{[2]^2} \right|^2 = \frac{(1+2\cos(2x)+\cos(4nx))^2+\sin(4nx)^2}{(2\cos(x))^4}
\end{equation}
assuming $q=\exp(i x)$ and $x=2\pi/(2+k)$.

Since Fibonacci anyons and Ising anyons are associated with $SU(2)_3$ and $SU(2)_2$ Chern-Simons theories correspondingly \cite{Field2018},\cite{Simon2023hdq}, and braiding operators for them differ from the $k>3$ case, we will provide calculations for them separately.

\subsection{Ising anyons}
For Ising anyons Racah matrices and R-matrices eigenvalues are well-known. See for example \cite{Simon2023hdq}. So the braiding operator is equal to
\begin{equation}
\hat{\sigma} = \frac{1}{2}
\begin{pmatrix}
1 & 1 \\
1 & -1
\end{pmatrix}
\begin{pmatrix}
e^{-i\pi/8} & 0 \\
0 & ie^{-i\pi/8}
\end{pmatrix}
\begin{pmatrix}
1 & 1 \\
1 & -1
\end{pmatrix}
=
\frac{e^{i \pi/8}}{\sqrt{2}}
\begin{pmatrix}
1 & -i \\
-i & 1
\end{pmatrix}.
\end{equation}
Consequently, the turnaround operator for Ising anyons is equal to
\begin{equation}
\hat{\mathcal{T}}=\hat{\sigma}^2 = e^{i \pi/4}
\begin{pmatrix}
0 & -i \\
-i & 0
\end{pmatrix}.
\end{equation}
And the $\mathbb{P}_n$ probability is equal to
\begin{equation}
\mathbb{P}_n=\left|\cos\frac{\pi n}{2} \right|.
\end{equation}

\subsection{Fibonacci anyons}
As well as Ising anyons the Fibonacci anyons are well-known too. See for example \cite{101143PTPS176384}, \cite{Hadjiivanov2024jho} or \cite{Simon2023hdq}. So the braiding operator is equal to
\begin{equation}
\hat{\sigma} = f^{-2}
\begin{pmatrix}
1 & \sqrt{f} \\
\sqrt{f} & -1
\end{pmatrix}
\begin{pmatrix}
e^{-4i\pi/5} & 0\\
0 & e^{3i\pi/5}
\end{pmatrix}
\begin{pmatrix}
1 & \sqrt{f} \\
\sqrt{f} & -1
\end{pmatrix}
=
f^{-1}
\begin{pmatrix}
e^{4i\pi/5} & \sqrt{f}e^{-3i\pi/5} \\
\sqrt{f}e^{-3i\pi/5} & -1
\end{pmatrix},
\end{equation}
where $f=(\sqrt{5}+1)/2$. Also turnaround operator is equal to
\begin{equation}
\hat{\mathcal{T}}=\hat{\sigma}^{2} = f^{-2}
\begin{pmatrix}
1 & \sqrt{f} \\
\sqrt{f} & -1
\end{pmatrix}
\begin{pmatrix}
e^{2i\pi/5} & 0\\
0 & e^{-4i\pi/5}
\end{pmatrix}
\begin{pmatrix}
1 & \sqrt{f} \\
\sqrt{f} & -1
\end{pmatrix}
\end{equation}
and $n$ turnaround operator is equal to
\begin{equation}
\hat{\mathcal{T}}^n= f^{-2}
\begin{pmatrix}
1 & \sqrt{f} \\
\sqrt{f} & -1
\end{pmatrix}
\begin{pmatrix}
e^{2ni\pi/5} & 0\\
0 & e^{-4ni\pi/5}
\end{pmatrix}
\begin{pmatrix}
1 & \sqrt{f} \\
\sqrt{f} & -1
\end{pmatrix}.
\end{equation}
This leads to the probability
\begin{equation}
\mathbb{P}_n=\left|\frac{e^{2ni\pi/5}}{f^2} + \frac{e^{-4ni\pi/5}}{f} \right|^2.
\end{equation}

\section{$k$ retrieving algorithm}
\begin{wrapfigure}[18]{r}{0.47\textwidth}
    \vspace{-1.5cm}
    \centering
    \begin{tikzpicture}[decoration=brace, scale=0.9]
        \node at (0,0) {\includegraphics[scale=0.58]{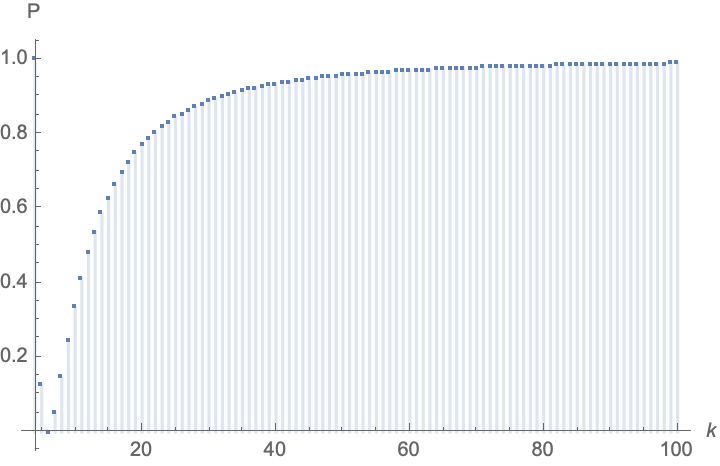}};
        \draw[decorate,thick] (-3.5,2.5)--(-2.4,2.5) node[midway, above]{hill};
        \draw[ultra thin] (-2.3,2.8)--(-2.3,-3);
        \draw[decorate,thick] (-2.2,2.5)--(3.7,2.5) node[midway, above]{tail};
        \node[fill=white] at (-3.9,2.5) {$\mathbb{P}_1$};
        \node[fill=white] at (4.1,-2.4){$k$};
        \draw[ultra thin] (-3.7,1.2)--(3.7,1.2);
    \end{tikzpicture}
    \caption{$\mathbb{P}_1(k)$, divided by hill and tail part}
    \label{fig:P1HillTail}
\end{wrapfigure}
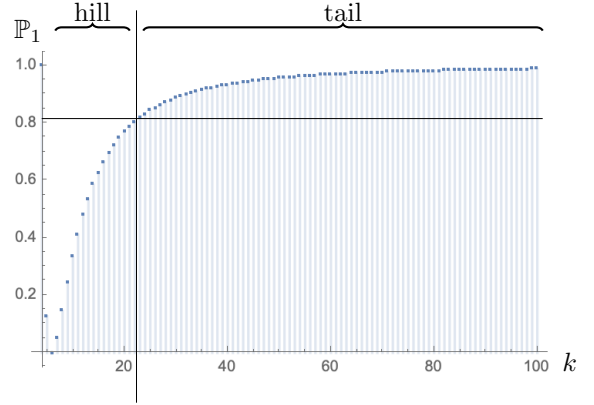
Here we will describe why one should start with $n=1$ turnarounds and take the next one from the list $(1,3,9,27,...,3^\#)$ every time when one is sure that the measured $\mathbb{P}_n$ is higher than $\mathbb{P}_n(k=24n-2) \approx 0.8$.
\subsection{Straightforward attempt problem}
If one wants to use only one turnaround during experiment, it can cause difficulties. Retrieving $k=160$ would require many more measurements than $k=10$, as retrieving $k=160$ would require a confidence interval being less or equal to $0.0001$, while extracting $k=10$ would require a confidence interval being less or equal to $0.168$. This happens because an infinite number of probabilities $\mathbb{P}_n(k)$ accumulate near 1. Let us divide the graph into a convenient and an inconvenient part for measurements. We call convenient for calculations part "hill" and inconvenient - "tail". The division as for now is approximately at $0.8$. So it is convenient to retrieve $k$ up to $22$. But what one should do to retrieve higher $k$?

\subsection{Using more turnarounds}
\begin{wrapfigure}[10]{r}{0.47\textwidth}
\vspace{-4cm}
    \centering
    \begin{tikzpicture}[scale = 0.9, decoration=brace]
        \node at (0,0) {\includegraphics[scale=0.58]{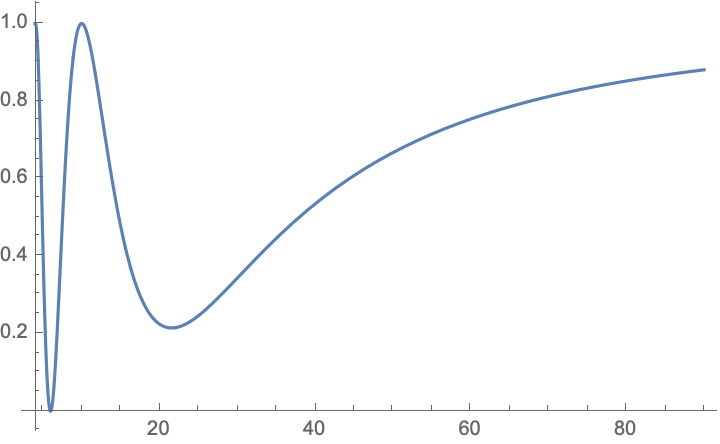}};
        \draw[decorate,thick] (-3.6,3.5)--(-2.2,3.5) node[midway, above, align=center]{oscillating \\ part};
        \draw[ultra thin] (-2.15,3.8)--(-2.15,-3);
        \draw[ultra thin] (2.15,3)--(2.15,-3);
        \draw[decorate,thick] (-2.1,3.5)--(3.9,3.5) node[midway, above]{monotonous part};
        \draw[decorate,thick] (-2.1,2.6)--(2.1,2.6) node[midway, above]{hill};
        \draw[decorate,thick] (2.2,2.6)--(3.9,2.6) node[midway, above]{tail};
        \node[fill=white] at (-3.6,2.7) {$\mathbb{P}_3$};
        \node[fill=white] at (4.1,-2.4){$k$};
        \node[fill=white] at (-1.75,-1.7){$k=8n-2$};
        \node[fill=white] at (2.65,-1.7){$k=24n-2$};
        \draw[ultra thin] (-3.7,1.43)--(4,1.43);
    \end{tikzpicture}
    \caption{General view of $\mathbb{P}_n$ \\ (here is $\mathbb{P}_3$ example)}
    \label{fig:plotGenView}
\end{wrapfigure}
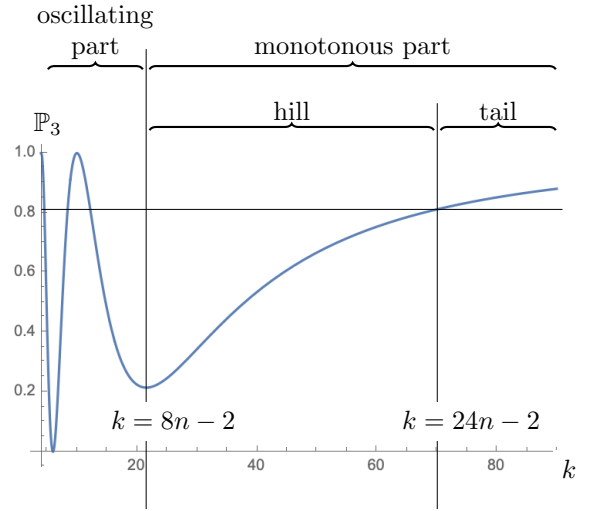
Take a look at equation \eqref{eq:main} one more time:
\begin{equation}
    \mathbb{P}_n(k) = \frac{(1+2\cos(2x)+\cos(4nx))^2+\sin(4nx)^2}{(2\cos(x))^4},
\end{equation}
where $x=2\pi/(2+k)$. Let us mention that $\mathbb{P}_n(k)\xrightarrow{k\to\infty} 1$, thus we get $\mathbb{P}_n(k)$ accumulating near 1. Also, the rightest minimum is at $k=8n-2$. We call the part before $k=8n-2$ the oscillating part. After one is sure $\mathbb{P}_1$ is higher than 0.8 ($k$ is higher than 22) one should try to use more turnarounds. Using $n=2$ turnarounds is not so effective, because its hill part starts at $k=8*2-2=14$. Thus hill part for $n=2$ turnaround covers $k$ from 14 to 22 which were already ruled out by $n=1$. For $n=3$ turnarounds this issue no longer stays true, since the hill part starts at $k=8*3-2=22$. Finally it is convenient to define hill part for $n$ turnarounds starting at the last minimum $k=8n-2$ and ending at $k=8*3n-2=24n-2$ and take the next $k$ from the following list $(1,3,9,27,...,3^\#)$. Thus the hill parts cover all possible $k$ and do not overlap with each other. For the endpoint $\mathbb{P}_n(k=24n-2)\approx0.8$.

\subsection{$k=160$ example}
\begin{figure}[h!]
    \centering
    \begin{tikzpicture}
    \node at (-10,0) {\includegraphics[scale=0.6]{Untitled-2.png}};
    \fill[red!50, semitransparent] (-13.45,1.3) rectangle (-6.6,1.9); 
        \node[fill=white] at (-13.5,2.5) {$\mathbb{P}_1$};
        \node[fill=white] at (-6.2,-2){$k$};
        \draw[->,thick] (-5.5,0)--(-4.5,0)node[midway,above]{1};
        \node at (0,0) {\includegraphics[scale=0.6]{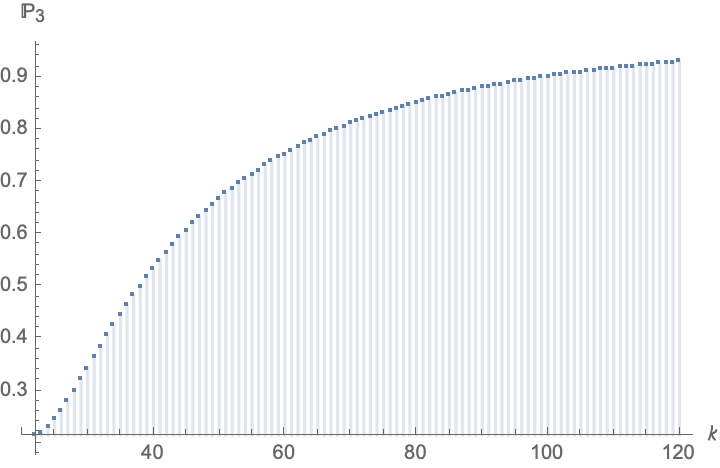}};
        \node[fill=white] at (-3.5,2.5) {$\mathbb{P}_3$};
        \node[fill=white] at (3.9,-2){$k$};
         \fill[red!50, semitransparent] (-3.45,1.3) rectangle (3.4,2.2); 
        \node at (0,-5.8) {\includegraphics[scale=0.6]{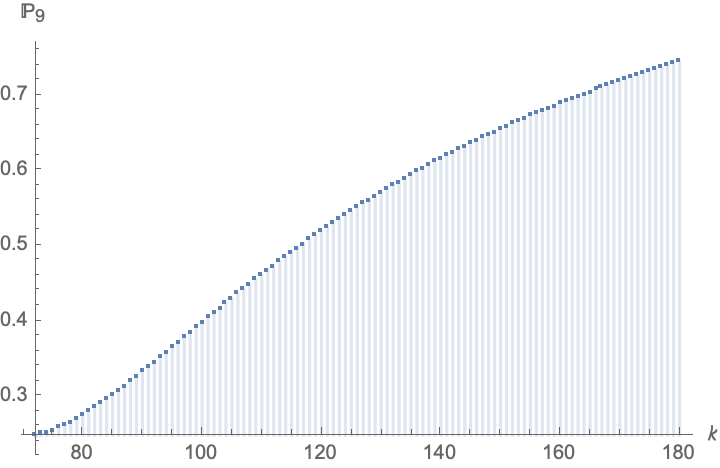}};
        \node[fill=white] at (-3.5,-3.4) {$\mathbb{P}_9$};
        \node[fill=white] at (3.9,-7.8){$k$};
        \fill[red!50, semitransparent] (-3.45,-4.3) rectangle (3.4,-5); 
        \node at (-10,-5.8) {\includegraphics[scale=0.6]{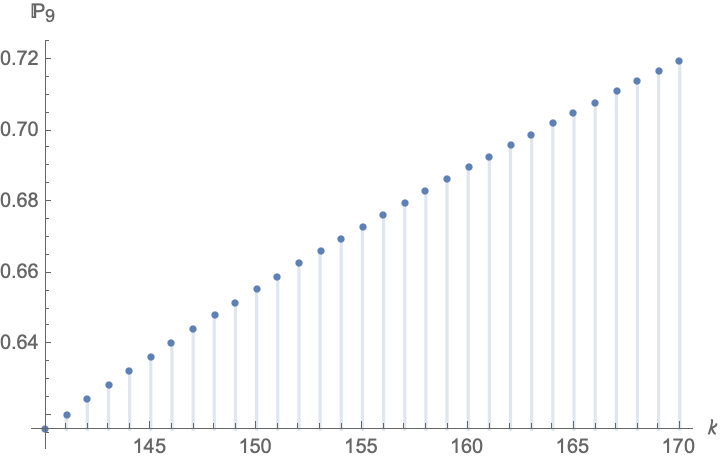}};
        \draw[->,thick] (-4.5,-7)--(-5.5,-7)node[midway,above]{3};
        \node[fill=white] at (-6.2,-7.8){$k$};
        \node[fill=white] at (-13.4,-3.4) {$\mathbb{P}_9$};
        \fill[red!50, semitransparent] (-13.35,-5.05) rectangle (-6.6,-5.25); 
        \draw[->,thick] (0,-2.7)--(0,-3.7)node[midway,right]{2};
\end{tikzpicture}
    \caption{$k=160$ retrieving example}
    \label{fig:k160}
\end{figure}

Let us now describe $k=160$ retrieving example (fig. \ref{fig:k160}). One starts with $n=1$ turnaround and provides measurements until $\mathbb{P}_1$ is definitely (taking into account confidence interval) above $\mathbb{P}_1(22)\approx 0.8$ (red line on fig. \ref{fig:k160} is the confidence interval). Then go to the $n=3$ measurements and repeat. Finally, with $n=9$ turnarounds $\mathbb{P}_9$ is definitely (again, taking into account confidence interval) below $\approx 0.8$. Here, the experimenter has to repeat measurements until the confidence interval covers the only $\mathbb{P}_9(k=160)$. 

\subsection{Possible experiment or algorithm improvements}
\subsubsection{Using oscillating part}
\begin{figure}[h!]
    \centering
    \begin{tikzpicture}
        \node at (0,-7) {\includegraphics[scale=0.55]{picpic9.png}};
        \node[fill=white] at (-3.3,-4.7) {$\mathbb{P}_9$};
        \node[fill=white] at (3.5,-9){$k$};
        \fill[red!50, semitransparent] (-3.15,-5.6) rectangle (3.1,-6.3); 
        \node at (-10,-7) {\includegraphics[scale=0.55]{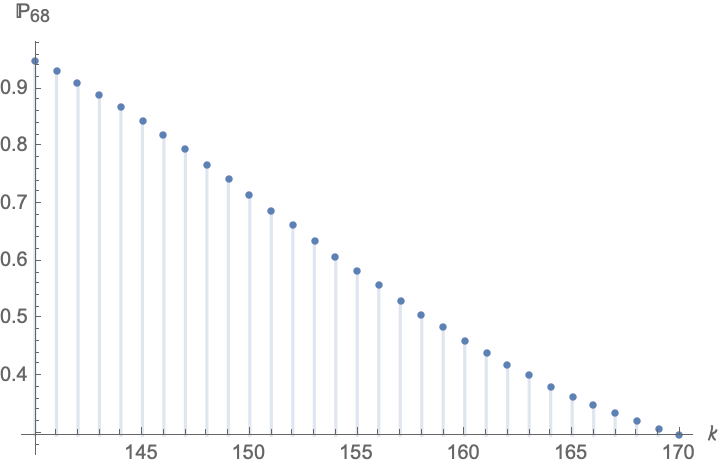}};
        \draw[->,thick] (-4.5,-7)--(-5.5,-7)node[midway,above]{3};
        \node[fill=white] at (-6.5,-9){$k$};
        \node[fill=white] at (-13.3,-4.7) {$\mathbb{P}_{68}$};
        \fill[red!50, semitransparent] (-13.15,-7.9) rectangle (-6.7,-8.15); 
\end{tikzpicture}
    \caption{$k=160$ retrieving using oscillating part}
    \label{fig:k160osc}
\end{figure}
Proposed algorithm can possibly be improved further. For example one can find it is sometimes convenient to use oscillating part while retrieving $k$. For $k=160$ example this means that after one starts to use $n=9$ turnarounds and realises that $\mathbb{P}_9$ is definitely below $\approx 0.8$, one can go to the $n>9$ ($n=68$ here) since it will require fewer measurements to get confidence interval small enough to retrieve k.

This is a fairly simple improvement to the algorithm, but it requires additional computation. It is necessary to specify at what confidence interval to switch to the oscillation part and how many turnarounds to use. Since this is beyond the main point of this paper, we do not provide further considerations on this matter.

\subsubsection{Measurable $\ydiagram{1}\otimes\ydiagram{1} \to \ydiagram{2}$ fusion}
If the experimentally obtained anyons are arranged in such a way that the $\ydiagram{1}\otimes\ydiagram{1} \to \ydiagram{2}$ fusion will mean not only the pair of anyons being in the $\ydiagram{2}$ representation, but also their fusion to another particle which is physically manifested and/or can be measurable. Then there is no need to measure the probability of annihilation $p(d,t)$. Then after $n$ turnarounds the experimenter is asked to wait for annihilation or measurable fusion. Annihilation will mean that the anyons were in representation $\varnothing$, and fusion, that they were in representation $\ydiagram{2}$ .

\subsubsection{Two pairs of anyons}
\begin{wrapfigure}[10]{r}{0.27\textwidth}
\vspace{-0.5cm}
    \centering
    \begin{tikzpicture}[scale=0.7, decoration=brace]
        \draw [ultra thick, ->] (-1,0) -- (-1,4)node[left]{Time\;};
        \newanyon(1,1);
        \newanyon(3,1);
        \rcross(2,1);
        \rcross(2,2);
        \draw[blue, thick](1,1)--(1,4);
        \draw[blue, thick](4,1)--(4,3.5);
        \draw[blue, thick](2,3)--(2,4);
        \draw[blue, thick](3,3)--(3,3.5);
        \draw[decorate,thick] (2.95,3.6)--(4.05,3.6) node[midway, above, align=center]{M};
    \end{tikzpicture}
    \caption{}
    \label{fig:impr}
\end{wrapfigure}
In section \ref{sec:whatabout} was stated that one has to somehow obtain an additional anyon (the first one on fig. \ref{fig:desc2}) in advance. How can this additional anyon be obtained? This anyons could be a part of another anyon pair produced from vacuum (fig. \ref{fig:impr}). It may seem that in this way one can get two measurements (M) instead of one after $n$ turnarounds (measuring left and right pair). However, this is not true. As we have mentioned before, the turnaround operator $\hat{\mathcal{T}}^n$ does not change the representation of the whole system $\ydiagram{1}\otimes\ydiagram{1}\otimes\ydiagram{1}\otimes\ydiagram{1}$. This means that if the right anyon pair is in the $\varnothing$ representation after the measurement, then the left anyon pair also must be in $\varnothing$ representation. Analogously, if the right anyon pair is in $\ydiagram{2}$ representation, then the left one also must be in $\ydiagram{2}$ representation to make the $\ydiagram{2}\otimes\ydiagram{2}\to\varnothing$ fusion possible. Therefore, the result of two measurements are correlated.

\subsection{Possible issues}
\subsubsection{$k=4$ problem}
Since 
\begin{equation}
\mathbb{P}_n(4)=1, 
\end{equation}
it becomes difficult to distinguish between $k=4$ and $k\to \infty$. Therefore, while making measurements, at one point the experimenter will have to stop and say that k is too large. However, since it is inefficient to build topological quantum computer algorithms at both k=4 and too large k, therefore these are models of little interest to quantum computations.

\subsubsection{Turning around more then one anyon}
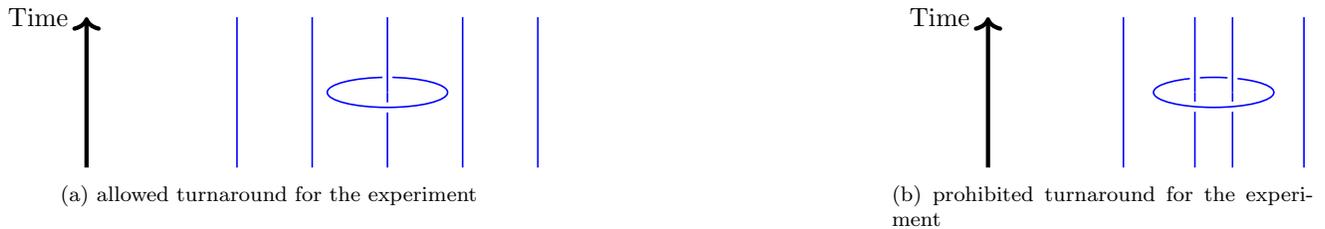
\begin{figure}[h!]
  \begin{center}
    \subfloat[allowed turnaround for the experiment]{
    \begin{tikzpicture}
    \draw [ultra thick, ->] (-4,-1) -- (-4,1)node[left]{Time\;};
    \draw[draw=blue, semithick] (-2,-1)--(-2,1);
    \draw[draw=blue, semithick] (-1,-1)--(-1,1);
    \draw [draw=white, double=blue, ultra thick](0,0)--(0,-1);
    \draw [draw=white, double=blue, ultra thick] (0,0) circle [ x radius=8mm, y radius=2mm];
    \draw [draw=white,double=blue,ultra thick] (0,0)--(0,1);
    \draw[draw=blue, semithick] (2,-1)--(2,1);
    \draw[draw=blue, semithick] (1,-1)--(1,1);
    \end{tikzpicture}
    }
    \hfill
    \subfloat[prohibited turnaround for the experiment]{
    \begin{tikzpicture}
    \draw [ultra thick, ->] (-3,-1) -- (-3,1)node[left]{Time\;};
    \draw[draw=blue, semithick] (-1.2,-1)--(-1.2,1);
    \draw[draw=blue, semithick] (1.2,-1)--(1.2,1);
    \draw [draw=white, double=blue, ultra thick](0.25,0)--(0.25,-1);
    \draw [draw=white, double=blue, ultra thick](-0.25,0)--(-0.25,-1);
    \draw [draw=white, double=blue, ultra thick] (0,0) circle [ x radius=8mm, y radius=2mm];
    \draw [draw=white,double=blue,ultra thick] (0.25,0)--(0.25,1);
    \draw [draw=white,double=blue,ultra thick] (-0.25,0)--(-0.25,1);
    \end{tikzpicture}
    \label{fig:prohib}
    }
  \end{center}
  \caption{Turn around more than one anyon is prohibited because superposition of two or more anyons can be in different states.}
\end{figure}

One should not wrap anyons around more than one anyon. Since two or more anyons can be in some superposition of states, after such a turnaround the probability of annihilation will depend on the state of anyons. If the experimenter knows the exact state of the group of anyons (fig. \ref{fig:prohib}) around which the turnover takes place, the calculations become more complicated, but still possible. However operators would be differ from $\hat{\mathcal{T}}^n$ (derived in sec. \ref{sec:alg}), and should be derived separately. If the experimenter does not know the states of the group of anyons around which the turnover occurs, then the calculations become impossible, since now the probability of annihilation depends not only on k, but also on the state of the group of anyons.

Let us also note that due to topological nature of anyons and
calculations, the presence of other anyons in the material does not affect the result of the calculations as long as they do not form nodes or meshes with the world lines of the anyons involved in the measurement.

\section{Conclusion}
Quantum gates depending on the Chern-Simons level  $k$ were derived earlier. Thus, having material in which $k$ is known in advance, the experimenter could construct the necessary algorithms using braid operators. However, the inverse problem of computing k in new (unknown) material has not been clarified. In this paper we suggested an algorithm for computing Chern-Simons level $k$ when the material contains $SU(2)_k$ Chern-Simons anyons, which are the most studied types of anyons. We also note some actions and caveats that may obstruct or complicate the computation.

This algorithm can be improved, and some suggestions were listed in this paper. However this is a subject of further studies. Also it is interesting to generalize this approach to other types of anyons, such as $SU(N)_k$ Chern-Simons anyons. This should be rather straightforward generalization from the algebraic point of view. However analysis of the possible results and their interpretation way be not as simple, especially since it is reasonable to try to find both $N$ and $k$ parameters.

\section*{Acknowledgments}
We would like to express our sincere appreciation to the Alexey Morozov, whose contributions and support have greatly enhanced the quality and rigour of this research. We are also grateful to Nikita Kolganov for helpful discussions and to the LMTPh scientific group for their attention during the presentations.

This work was supported by the Russian Science Foundation grant No 23-71-10058.
\newpage
\printbibliography
\end{document}